# fastMRI Breast: A publicly available radial k-space dataset of breast dynamic contrast-enhanced MRI

[1,2]Eddy Solomon, [2]Patricia M. Johnson, [3]Zhengguo Tan, [2]Radhika Tibrewala, [2]Yvonne W. Lui, [2,3]Florian Knoll, [2]Linda Moy, [1,2]Sungheon Gene Kim, [2]Laura Heacock

[1]Department of Radiology, Weill Cornell Medical College, New York, NY, USA.
[2]Center for Advanced Imaging Innovation and Research (CAI2R), Department of Radiology, New York University Grossman School of Medicine, New York, NY, USA.
[3]Department Artificial Intelligence in Biomedical Engineering, University Erlangen-Nuremberg, Erlangen, Germany.

Corresponding author: Eddy Solomon, Ph.D. eds4001@med.cornell.edu

## Summary:

This data curation work introduces the first large-scale dataset of radial k-space and DICOM data for breast DCE-MRI acquired in diagnostic breast MRI exams. Our dataset includes case-level labels indicating patient age, menopause status, lesion status (negative, benign, and malignant), and lesion type for each case. The public availability of this dataset and accompanying reconstruction code will support research and development of fast and quantitative breast image reconstruction and machine learning methods.

## Introduction:

Breast DCE-MRI is widely used for three main indications: screening patients at high-risk for breast cancer (1), staging (2), and predicting response to chemotherapy (3, 4). While current breast DCE-MRI exams mainly depend on qualitative evaluation of morphological features before and after contrast, the latest acquisition protocols attempt to integrate both detailed morphological information, derived from high spatial resolution images, and contrast kinetic information (5). However, this approach poses significant challenges in image reconstruction due to the inherent trade-off between temporal resolution and image quality. To address this challenge, acquisition of golden-angle radial DCE-MRI has been proposed combined with compressed sensing and parallel imaging, offering high spatial resolution and flexible temporal resolution, while maintaining comparable diagnostic performance to conventional DCE-MRI (6). Recently, there have also been explorations into the development of deep learning reconstruction of DCE-MRI data (7). Ultimately, with the latter, we can improve both morphological feature evaluation and extraction of contrast kinetic information to improve breast cancer diagnosis. However, there has been no publicly available dataset that can be utilized for these developments and for comparison of different reconstruction methods.

Most datasets (8-10) are limited to reconstructed images and do not include raw k-space data. To fulfil this shortcoming, the fastMRI initiative has recently been taken to make the raw k-space data of the knee (11), brain (12) and prostate (13) publicly available. Nevertheless, so far, publicly available MRI datasets have been restricted to Cartesian sampling sequences and did not include any dynamic contrast-enhanced MRI data. Consequently, the lack of large open-source databases for DCE-MRI has posed a significant hurdle to the development of machine learning and quantitative image reconstruction techniques in this field. To address this unmet need, here we present a first release of a radial acquisition dataset of breast DCE-MRI. The dataset includes breast MRI with case-level labels, where each case includes raw radial k-space data and a corresponding example of reconstructed image series and Digital Imaging and Communications in Medicine (DICOM) files. Finally, to improve user accessibility and reproducibility, our dataset is accompanied with open-source image reconstruction code.

## Description of dataset

### Patient population:

The collection and sharing of this dataset were approved by our institutional review board and was compliant with Health Insurance Portability and Accountability. Given the retrospective nature of this data collection, it received a waiver of consent from our institutional review board. The dataset includes 300 patients with and without breast cancer, acquired as a part of clinical breast MRI exams between Dec 2019 and June 2022. 90 cases were found to have at least one malignant lesion, 159 benign and 51 cases had no suspicious lesions. Malignant lesions include invasive ductal carcinomas (IDC) (n=56), invasive lobular carcinomas (ILC) (n=10), a mix of IDC and ILC (n=6) and ductal carcinomas in-situ (DCIS) (n=18). Among the 300 cases, sixteen patients had a repeated breast MRI after a year. The mean age of subjects is 44±12 years. Our dataset also includes a case-level labels describing, for each case, the patient age, menopause status, lesion status (negative, benign and malignant) and malignant lesion type. This study included also routine clinical Cartesian sequences, however, here we release only the non-Cartesian datasets with contrast injection.

### Breast DCE-MRI study:

All subjects underwent a diagnostic breast MRI exam on a whole-body 3T scanner (MAGNETOM TimTrio, Siemens Healthcare) with a 16-channel breast coil. The DCE-MRI was conducted with a free-breathing golden-angle radial VIBE sequence and reconstructed with Golden-angle RAdial Sparse Parallel (GRASP) (14). A total of 288 spokes were acquired continuously for 2.5 min with 83 partitions and partial Fourier of 6/8, resulting in 192 slices (see scan parameters in Table 1). A GRASP scan was first acquired before the administration of contrast (Seg 1) and a second GRASP scan was acquired with contrast injection (Seg 2) (Fig. 1a). The radial trajectories of these two scans are same; the radial angle of each scan started at zero degree. For the second scan, a single dose of Gadobutrol (Gadavist, Bayer Healthcare Pharmaceuticals) at 0.1 mM/kg body weight was injected at 2 mL/s intravenously, after 30-seconds into the scan. Both scans can be used for the development of static and dynamic image reconstruction and machine learning methods, while the second one alone for the development of quantitative breast DCE-MRI methods (15).

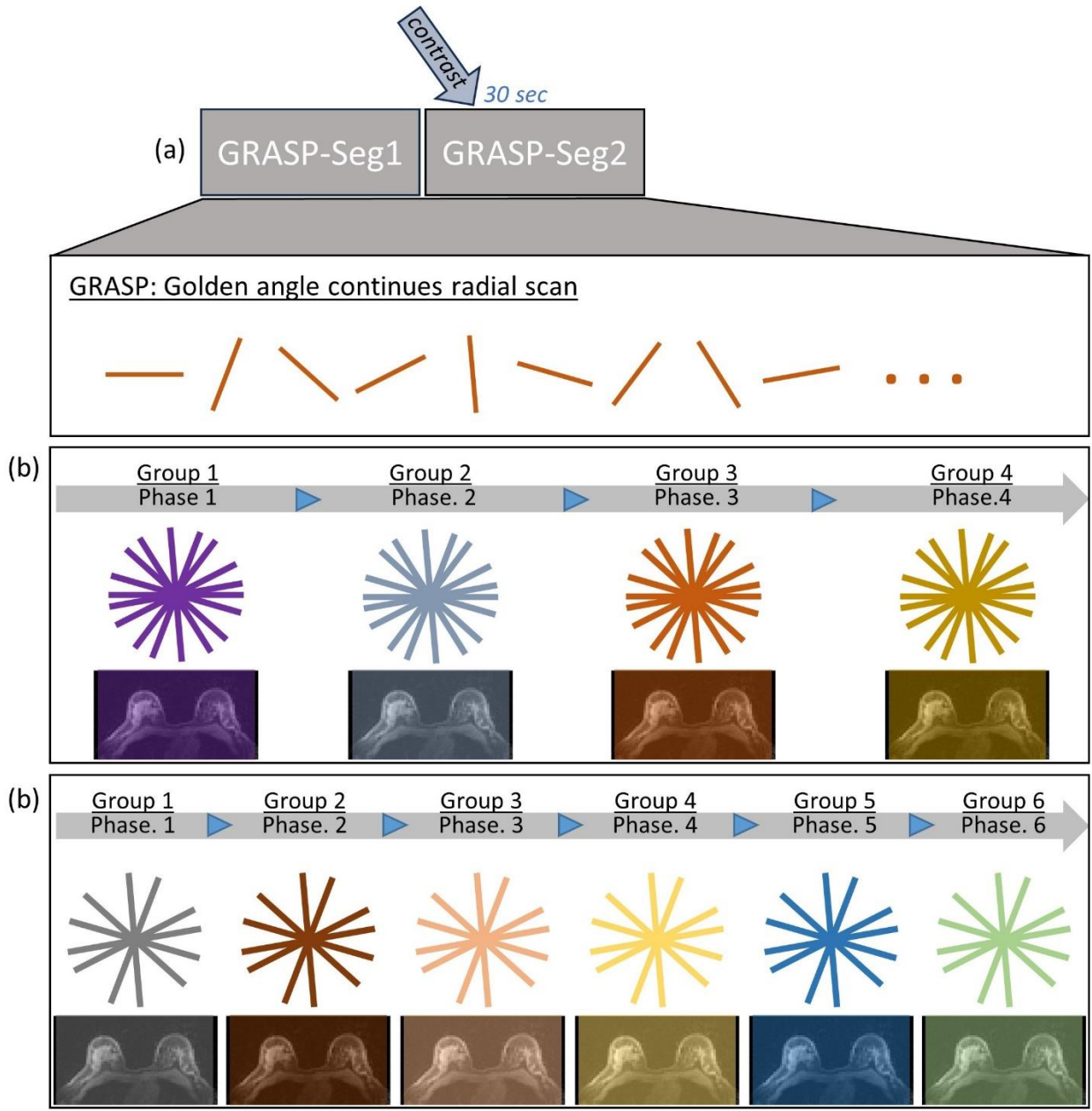

**Figure. 1:** Breast DCE-MRI study design. (a) GRASP DCE-MRI scan of 2.5 min was acquired once before the administration of contrast (Seg1) and a second scan was acquired with contrast injection (Seg2). Using the reconstruction Python code, (b, c) one can generate undersampled GRASP image series with flexible temporal resolutions. DCE: Dynamic Contrast-Enhanced GRASP: Golden-angle RAdial Sparse Parallel imaging.

**Raw k-space dataset:**

Raw k-space data were exported from the scanners, anonymized, and saved into a Hierarchical Data Format (HDF5) file. Each radial raw k-space dataset was saved into a separate h5 data file with relevant acquisition parameters, and a representative 4D image series (x, y, z, time) reconstructed with 4-time frames. To enhance accessibility, image series were also saved in DICOM format. The files of each scan are saved under a folder named with a coded scan number and segment number as suffix (e.g., 'fastMRI_breast_133_1'). The size of each case folder is ~ 4.5 GB. To promote the use of this dataset for future development of machine learning tools and their comparisons, patient cases are split into training and testing cohorts at 80:20 ratio, respectively (Table 2), while also keeping the same ratio per lesion type as closely as possible. The data are available for free through: https://fastmri.med.nyu.edu/. After acceptance of the dataset sharing agreement, researchers are provided a link to the download webpage. Clinical labeling information, data split and repeated scan identifiers are provided in a separate excel sheet also available at the provided link.

**Reconstruction code:**

We provide an example GPU-based GRASP reconstruction code implemented in Python together with this dataset. The code can be downloaded through: https://github.com/eddysolo/demo_dce_recon. It is based on a SigPy (16) implementation of non-uniform fast Fourier transform (NUFFT) (17), coil sensitivity estimation(18) and iterative reconstruction with temporal total variation (TV) regularization. This code demonstrates how our radial k-space data can be used to reconstruct dynamic images with a flexible temporal resolution (i.e., number of radial views per frame) (Fig. 1b and c), which are also saved as HDF5 and DICOM files. Representative reconstructed images of malignant, benign and healthy datasets are shown in Fig. 2a. In the presence of a cancer lesion (Fig. 2b), one can appreciate rapid enhancement following contrast injection reconstructed with various temporal resolutions using the provided code (Fig. 2c).

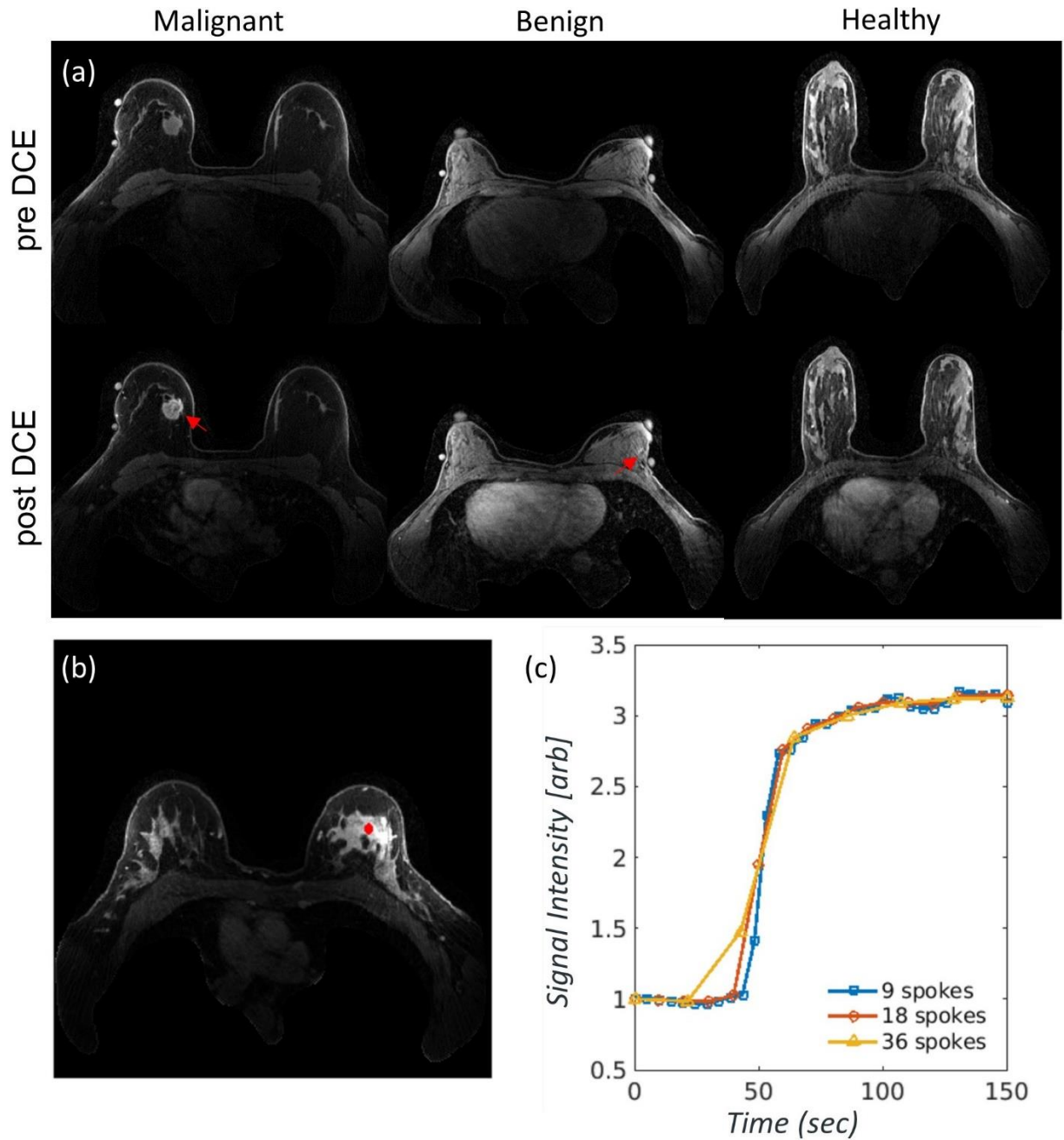

**Figure. 2:** Representative examples of Breast DCE-MRI reconstruction. (a) A malignant, benign and healthy dataset at pre and post contrast injection, with 5.3 sec temporal resolution. (b) A malignant lesion and its (c) signal enhancement during injection of contrast material, reconstructed with temporal resolution of 4.7 sec (9 spokes), 9.4 sec (18 spokes) and 18.8 sec (36 spokes).

## Discussion:

To our knowledge, this dataset represents the first radial k-space dataset of breast DCE-MRI derived from a clinical cohort, openly accessible to the public. While public breast DCE datasets do exist, they are generally focused on particular research questions with already reconstructed images, such as segmentation and classification (19) or machine learning performance (20). Consistent with increasing awareness of the importance of medical data sharing (21, 22), there is a growing recognition of the advantages of providing free access to raw data, rather than reconstructed images. Thus, by utilizing our k-space data, biased results in model performance due to hidden processing pipelines can be avoided (23). Moreover, unlike machine-learning models trained exclusively on datasets from healthy individuals or with limited pathology cases, this dataset includes a variety of malignant and benign cases in both training and testing cohorts. This diversity can enhance the generalizability of future DCE-MRI models. Finally, utilizing this dataset alongside an open-source code can further enhance the reproducibility of future research results.

## Acknowledgements:

RSNA Research Seed Grant RSD1830, NIH P41EB017183, NIH R01CA160620, R01CA219964, and UH3CA228699

**Tables:**

| Table1: Breast data scanning parameters | |
|---|---|
| Sequence | T1 weighted radial stack-of-stars 3D gradient echo |
| FOV | 320 × 320 × 212 mm$^3$ |
| Matrix size | 320 × 320 |
| Slice Thickness | 1.1mm |
| TR | 4.87ms |
| TE | 1.8ms |
| Number of slices | 192 |
| Radial views | 288 |
| Spatial resolution | 1 × 1 × 1.1mm$^3$ |
| Fat suppression | Spectrally selective Adiabatic Inversion Recovery (SPAIR) |

## Table 2: Dataset Split

Malignant lesions: Ductal carcinoma in situ (DCIS), Invasive ductal carcinoma (IDC), Invasive lobular carcinoma (ILC)

| 300 cases (80:20 ratio) | Training (240 cases) | Testing (60 cases) |
|---|---|---|
| **Negative** | 41 | 10 |
| **Benign** | 127 | 32 |
| **Malignancy** | 72 | 18 |
| DCIS | 14 | 4 |
| IDC | 46 | 10 |
| ILC | 8 | 2 |
| IDC/ILC | 4 | 2 |